\documentclass[12pt]{elsarticle}
\usepackage{slashed}
\usepackage{epsfig,graphics}
\usepackage{hyperref}
\usepackage{bbm}
\usepackage{mathtools}
\usepackage{amssymb}
\usepackage{mathrsfs}
\usepackage{amsmath}
\usepackage{natbib}
\usepackage{color}

\newlength{\llslash}

\newcommand{\tr}{\rm tr \,}

\def\tr{{\rm tr} \,}

\def\Dslash{D \hspace{-2.7mm}/ \;}

\def\w2{\tilde w^2}
\def\ws2{1}

\begin{document}

\begin{frontmatter}
\title{Low-energy constants \\from charmed baryons on QCD lattices}
\author[SUT]{Yonggoo Heo}
\author[GSI]{Xiao-Yu Guo}
\author[GSI,TUD]{M.F.M. Lutz}
\address[SUT]{Suranaree University of Technology, Nakhon Ratchasima, 30000, Thailand}
\address[GSI]{GSI Helmholtzzentrum f\"ur Schwerionenforschung GmbH,\\
Planck Str. 1, 64291 Darmstadt, Germany}
\address[TUD]{Technische Universit\"at Darmstadt, D-64289 Darmstadt, Germany}
\begin{abstract}
We study the light quark-mass dependence of charmed baryon masses as measured by various QCD lattice collaborations. A global fit to such data based on the chiral SU(3) Lagrangian is reported on. All low-energy constants that are relevant at  next-to-next-to-next-to-leading  order (N$^3$LO) are determined from the lattice data sets where constraints from sum rules as they follow from large-$N_c$ QCD at subleading order are  considered. The expected hierarchy  for the low-energy constants in the $1/N_c$ expansion is confirmed by our global fits to the lattice data. With our results the low-energy interaction of the Goldstone  bosons with the charmed baryon ground states is well constrained and the path towards realistic coupled-channel computations in this sector of QCD is prepared. 
\end{abstract}
\end{frontmatter}

\section{Introduction}

The challenge of modern particle physics is how to connect the wealth of experimental data on hadron spectroscopy to the fundamental theory of strong interactions. While lattice QCD approaches made significant advances in the last decade, it is still a grand endeavor  
to compute the hadronic excitation spectrum on QCD lattices directly. On the one hand the reliable determination of finite-box energy levels requires the consideration of all open  hadronic channels as it was demonstrated repeatedly that the omission of some channels  may distort the finite-box spectrum significantly \cite{Briceno:2017max}. On the other hand the translation of such finite-box spectra to the physics in the laboratory requires extensive coupled-channel studies that resemble to some extent the  partial-wave analyses required for the experimental data sets. 

Given the complexity of this problem it may be of advantage to systematically combine the 
strength of lattice QCD approaches with that of effective field theory (EFT). While an EFT 
is quite efficient to work in symmetry aspects of QCD and gets close to data taken in the laboratory it is a priori ignorant about various dynamical aspects of QCD.  This is reflected in the large number of low-energy constants that arise at subleading orders. However, we argue that this main drawback can be overcome  by the use of existing lattice data sets. Since the simulations are performed  at various unphysical quark masses information is generated that is instrumental in determining large sets of low-energy constants. 

Such programs have already been successfully  set up for masses of baryons and  mesons in their ground states with $J^P = \frac{1}{2}^+, \frac{3}{2}^+$ and 
$J^P = 0^-, 1^-$ quantum numbers \cite{Lutz:2018cqo,Guo:2018kno,Heo:2018qnk,Guo:2018zvl}. 
Significant sets of low-energy parameters to be used in flavour SU(3) chiral Lagrangians 
were established from the available lattice data on such hadron masses. So far results are available for mesons and baryons composed of up, down and strange quarks \cite{Lutz:2018cqo,Guo:2018zvl,Guo:2018pyq}. In addition a complete set of relevant low-energy constants in the open-charm sector of QCD was established in \cite{Guo:2018kno}. 

The purpose of this Letter is to present results for the low-energy constants of the chiral Lagrangian formulated for charmed baryons \cite{Yan:1992gz,Cho:1992gg,Lutz:2003jw,Lutz:2014jja}. Like in the open-charm meson sector the coupled-channel dynamics of the leading order chiral Lagrangian predicts the existence of various dynamically generated states \cite{Lutz:2003jw,Hofmann:2005sw,Hofmann:2006qx}. 
Such states carry $J^P =\frac{1}{2}^+$and $J^P=\frac{3}{2}^+$ quantum numbers and  may be formed in flavour exotic multiplets. Thus it is an important challenge to estimate the size 
of chiral correction terms from QCD studies. 
At N$^3$LO we count 54 parameters that need to be determined in computations of the charmed baryon masses. Given the limited data set  that is provided so far \cite{Liu:2009jc,Briceno:2012wt,Alexandrou:2012xk,Namekawa:2013vu,Brown:2014ena,Alexandrou:2014sha,Bali:2015lka} it is important to derive additional constraints from QCD that further constrain a fit of the low-energy constants to such data. A corresponding framework was worked out already in \cite{Heo:2018mur,Heo:2018qnk}.
From a systematic consideration of sum rules that arise in the $1/N_c$ expansion of QCD at subleading order the number of independent low-energy constants was reduced significantly down to 23 parameters only \cite{Heo:2018mur}.

\section{Chiral Lagrangian with charmed baryon fields} \label{section:chiral-lagrangian}

The chiral Lagrangian combined with appropriate counting rules leads to  systematic results in low-energy hadron physics. In the following we recall 
the leading order (LO) terms in the open-charm baryon sector of QCD \cite{Yan:1992gz,Cho:1992gg,Lutz:2014jja}. It is convenient to decompose the fields into their isospin multiplets with
\begin{eqnarray}
&& \Phi = \tau \cdot \pi (140)
+ \alpha^\dagger \cdot  K (494) +  K^\dagger(494) \cdot \alpha
+ \eta(547)\,\lambda_8\,,
\nonumber\\
&&  \sqrt{2}\,B_{[\bar 3]}  = {\textstyle{1\over \sqrt{2}}}\,\alpha^\dagger \cdot \Xi_c(2470)
- {\textstyle{1\over \sqrt{2}}}\,\Xi_c^{T}(2470)\cdot \alpha
+  i\,\tau_2\,\Lambda_c(2284) \,,
\nonumber\\
&& \sqrt{2}\,B_{[6]} = {\textstyle{1\over \sqrt{2}}}\,\alpha^\dagger \cdot \Xi'_c(2580)
+ {\textstyle{1\over \sqrt{2}}}\,{\Xi'}_{\!\!c}^{T}(2580)\cdot \alpha
+ \Sigma_c(2455) \cdot \tau \,i\,\tau_2 
\nonumber\\
&& \qquad \quad+  {\textstyle{\sqrt{2}\over 3}}\, \big(1-\sqrt{3}\,\lambda_8 \big)\,\Omega_c(2704)  \,,
\nonumber\\
&& \sqrt{2}\, B^\mu_{[6]} = {\textstyle{1\over \sqrt{2}}}\,\alpha^\dagger \cdot \Xi^{\mu}_c(2645)
+ {\textstyle{1\over \sqrt{2}}}\,\Xi_c^{T,\mu}(2645)\cdot \alpha
+ \Sigma^\mu_c(2520) \cdot \tau \,i\,\tau_2 
\nonumber\\
&& \qquad \quad+  {\textstyle{\sqrt{2}\over 3}}\, \big(1-\sqrt{3}\,\lambda_8 \big)\,\Omega^{\mu}_c(2770)  \,,
\nonumber\\
&& \alpha^\dagger = {\textstyle{1\over \sqrt{2}}}\,(\lambda_4+i\,\lambda_5 ,\lambda_6+i\,\lambda_7 )\,,\qquad
\tau = (\lambda_1,\lambda_2, \lambda_3)\,,
\end{eqnarray}
where the matrices $\lambda_i$ are the standard Gell-Mann generators of the SU(3) algebra.
The numbers in the brackets recall the approximate masses of the particles in units of
MeV. There are the kinetic terms 
\allowdisplaybreaks[1]
\begin{eqnarray}
&& \mathcal{L}^{(1)} \!= \tr \bar B_{[6]} \big(\gamma^\mu\, i\,D_\mu  - M^{(1/2)}_{[6]} \big)\,B_{[6]}
- \mathrm{tr}\,\Big(\bar{B}_{[6]}^\mu \, \big(\big[i\,\Dslash\,
-M^{(3/2)}_{[6]}\big]\,g_{\mu\nu} 
\nonumber\\
&& \qquad \quad \;
-\,i\,(\gamma_\mu\, D_\nu 
 + \gamma_\nu \,D_\mu) +\, \gamma_\mu\,\big[i\,\Dslash + M^{(3/2)}_{[6]}\big]\,\gamma_\nu \big)\,B_{[6]}^\nu\Big)
 \nonumber \\
&&  \qquad \;+\, \tr \bar B_{[\bar 3]} \big(\gamma^\mu\,i\, D_\mu  -M^{(1/2)}_{[\bar 3]} \big)\,B_{[\bar 3]} 
\nonumber \\
&&  \qquad \;+\,F_{[66]}\,{\rm tr}\,\bar B_{[6]}\,\gamma^\mu\,\gamma_5\,i\,U_\mu\,B_{[6]}
+ F_{[\bar 3\bar 3]}\,{\rm tr}\,\bar B_{[\bar 3]}\,\gamma^\mu\,\gamma_5\,i\,U_\mu\,B_{[\bar 3]}
\nonumber\\
&&  \qquad \;+\, F_{[\bar 36]}\,{\rm tr}\,\Big( \bar B_{[6]}\,\gamma^\mu\,\gamma_5\,i\,U_\mu\,B_{[\bar 3]} + {\rm h.c.}\Big)
\nonumber\\
&&  \qquad \;+\, C_{[66]}\,{\rm tr}\,\Big( \bar B_{[6]}^\mu\,i\,U_\mu\,B^{\phantom{\mu}}_{[6]} + {\rm h.c.} \Big)
 + C_{[\bar 36]}\,{\rm tr}\,\Big( \bar B_{[6]}^\mu\,i\,U_\mu\,B^{\phantom{\mu}}_{[\bar 3]} + {\rm h.c.}\Big)
\nonumber\\
&&  \qquad \;-\, H_{[66]}\,{\rm tr}\,\bar B_{[6]}^\alpha\,g_{\alpha \beta}\,\gamma^\mu\,\gamma_5\,i\,U_\mu\,B_{[6]}^\beta \,,
\nonumber\\
\nonumber\\
&& U_\mu = {\textstyle{1\over 2}}\,u^\dagger \, \big(
\partial_\mu \,e^{i\,\frac{\Phi}{f}} \big)\, u^\dagger
-{\textstyle{i\over 2}}\,u^\dagger \,(v_\mu+ a_\mu)\, u
+{\textstyle{i\over 2}}\,u \,(v_\mu-a_\mu)\, u^\dagger\;, 
\nonumber\\
&&  D_\mu \, B \;\,= \partial_\mu B +  \Gamma_{\mu}\, B + B\,\Gamma^T_{\mu} \,, \qquad \qquad \qquad  u = e^{i\,\frac{\Phi}{2\,f}}  \,,
\nonumber\\
&&\Gamma_\mu ={\textstyle{1\over 2}}\,u^\dagger \,\big[\partial_\mu -i\,(v_\mu + a_\mu) \big] \,u
+{\textstyle{1\over 2}}\, u \,
\big[\partial_\mu -i\,(v_\mu - a_\mu)\big] \,u^\dagger\,,  
\label{def-L1}
\end{eqnarray}
and 6 structures which parameterize the three-point interactions of the Goldstone bosons with the charmed baryon 
fields \cite{Yan:1992gz,Cho:1992gg}. At leading order in a chiral expansion, the bare masses 
$M^{(1/2)}_{[6]}$, $M^{(3/2)}_{[6]}$ and $M^{(1/2)}_{[\bar 3]}$ may be identified with the flavour average of the sextet and anti-triplet baryon masses.

At next-to-leading order (NLO) there are symmetry conserving 
and symmetry breaking terms \cite{Lutz:2014jja,Heo:2018mur}.  
A complete list of chiral symmetry conserving $Q^2$ counter terms was given in \cite{Lutz:2014jja,Heo:2018mur}. In these works the $Q^2$ counter terms are grouped according
to their Dirac structure. Here we display the scalar and vector terms relevant for our study only
\allowdisplaybreaks[1]
\begin{eqnarray}
&& \mathcal{L}^{(S)} \!= -g_{0,[\bar 3\bar 3]}^{(S)}\,{\rm tr}\,\big( \bar B_{[\bar 3]}\,B_{[\bar 3]}\big)\,{\rm tr}\,\big( U_\mu\,U^\mu\big)
- g_{D,[\bar 3\bar 3]}^{(S)}\,{\rm tr}\,\big( \bar B_{[\bar 3]}\,\big\{ U_\mu,\,U^\mu\big\}\, B_{[\bar 3]}\big)
\nonumber\\
&& \quad-\, 
g_{0,[66]}^{(S)}\,{\rm tr}\,\big( \bar B_{[6]}\,B_{[6]}\big)\,{\rm tr}\,\big( U_\mu\,U^\mu\big)
- g_{1,[66]}^{(S)}\,{\rm tr}\,\big( \bar B_{[6]}\, U^\mu\,B_{[6]}\, U^T_\mu\big)
\nonumber\\
&& \quad-\, g_{D,[66]}^{(S)}\,{\rm tr}\,\big( \bar B_{[6]}\,\big\{ U_\mu,\,U^\mu\big\}\, B_{[6]}\big)
- g_{D,[\bar 36]}^{(S)}\,{\rm tr}\,\big( \bar B_{[6]}\,\big\{ U_\mu,\,U^\mu\big\}\, B_{[\bar 3]} + {\rm h.c.}\big)
\nonumber\\
&& \quad+\, h_{0,[66]}^{(S)}\,{\rm tr}\,\big( \bar B_{[6]}^\mu\,g_{\mu \nu}\,B_{[6]}^\nu\big)\,{\rm tr}\,\big( U_\alpha\,U^\alpha\big)
+ h_{1,[66]}^{(S)}\,{\rm tr}\,\big( \bar B_{[6]}^\mu\,B_{[6]}^\nu\big)\,{\rm tr}\,\big( U_\mu\,U_\nu\big)
\nonumber\\
&& \quad+\, h_{2,[66]}^{(S)}\,{\rm tr}\,\big( \bar B_{[6]}^\mu\, g_{\mu \nu}\,\big\{ U^\alpha,\,U_\alpha\big\}\, B_{[6]}^\nu\big)
+ h_{3,[66]}^{(S)}\,{\rm tr}\,\big( \bar B_{[6]}^\mu\,\big\{ U_\mu,\,U_\nu\big\}\, B_{[6]}^\nu\big)
\nonumber\\
&& \quad+\, h_{4,[66]}^{(S)}\,{\rm tr}\,\big( \bar B_{[6]}^\mu\,g_{\mu \nu}\,U^\alpha\, B_{[6]}^\nu\,U^T_\alpha \big)
\nonumber\\
&& \quad+\, \frac{1}{2}\, h_{5,[66]}^{(S)}\,{\rm tr}\,\big( \bar B_{[6]}^\mu\,U_\nu\, B_{[6]}^\nu\,U^T_\mu + \bar B_{[6]}^\mu\,U_\mu\, B_{[6]}^\nu\,U^T_\nu\big)\,,
\nonumber\\
&& \mathcal{L}^{(V)} \!= -\frac{1}{2}\,g_{0,[\bar 3\bar 3]}^{(V)}\,{\rm tr}\,\big( \bar B_{[\bar 3]}\,i\,\gamma^\alpha\,(D^\beta B_{[\bar 3]})\,{\rm tr}\,\big( U_\beta\,U_\alpha\big) \big)
\nonumber\\
&& \quad-\, \frac{1}{2}\,g_{1,[\bar 3\bar 3]}^{(V)}\,{\rm tr}\,\big( \bar B_{[\bar 3]}\,i\,\gamma^\alpha\,U_\beta\,(D^\beta B_{[\bar 3]})\, U^T_\alpha + \bar B_{[\bar 3]}\,i\,\gamma^\alpha\,U_\alpha\,(D^\beta B_{[\bar 3]})\, U^T_\beta\big)
\nonumber\\
&& \quad-\,
 \frac{1}{2}\,g_{D,[\bar 3\bar 3]}^{(V)}\,{\rm tr}\,\big( \bar B_{[\bar 3]}\,i\,\gamma^\alpha\,\big\{ U_\alpha,\,U_\beta\big\}\,(D^\beta  B_{[\bar 3]}) \big)
\nonumber\\
&& \quad-\,  \frac{1}{2}\,g_{D,[\bar 36]}^{(V)}\,{\rm tr}\,\big( \bar B_{[6]}\,i\,\gamma^\alpha\,\big\{ U_\alpha,\,U_\beta\big\}\,(D^\beta  B_{[\bar 3]})
- (D^\beta \bar B_{[6]})\,i\,\gamma^\alpha\,\big\{ U_\alpha,\,U_\beta\big\}\, B_{[\bar 3]}
\big)
\nonumber\\
&& \quad-\,  \frac{1}{2}\,g_{0,[66]}^{(V)}\left(\,{\rm tr}\,\big( \bar B_{[6]}\,i\,\gamma^\alpha\,(D^\beta B_{[6]})\big)\,{\rm tr}\,\big( U_\beta\,U_\alpha\big) \right)
\nonumber\\
&& \quad-\, \frac{1}{4}\,g_{1,[66]}^{(V)}\,{\rm tr}\,\big( \bar B_{[6]}\,i\,\gamma^\alpha\,U_\beta\,(D^\beta B_{[6]})\, U^T_\alpha + \bar B_{[6]}\,i\,\gamma^\alpha\,U_\alpha\,(D^\beta B_{[6]})\, U^T_\beta\big)
\nonumber\\
&& \quad-\, \frac{1}{2}\,g_{D,[66]}^{(V)}\,{\rm tr}\,\big( \bar B_{[6]}\,i\,\gamma^\alpha\,\big\{ U_\alpha,\,U_\beta\big\}\,(D^\beta  B_{[6]})\big)
\nonumber\\
&& \quad+\,\frac{1}{2}\, h_{0,[66]}^{(V)}\,{\rm tr}\,\big( \bar B_{[6]}^\mu\,g_{\mu \nu}\,i\,\gamma^\alpha\,(D^\beta B_{[6]}^\nu)\,{\rm tr}\,\big( U_\alpha\,U_\beta\big) \big)
\nonumber\\
&& \quad+\,  \frac{1}{4}\,h_{1,[66]}^{(V)}\,{\rm tr}\,\big( \bar B_{[6]}^\mu\,g_{\mu \nu}\, i\,\gamma^\alpha\,U_\beta\,(D^\beta B_{[6]}^\nu)\, U^T_\alpha + \bar B_{[6]}^\mu\,g_{\mu \nu}\,i\,\gamma^\alpha\,U_\alpha\,(D^\beta B_{[6]}^\nu)\, U^T_\beta\big)
\nonumber\\
&& \quad+\, \frac{1}{2}\,h_{2,[66]}^{(V)}\,{\rm tr}\,\big( \bar B^\mu_{[6]}\,g_{\mu \nu}\,i\,\gamma^\alpha\,\big\{ U_\alpha,\,U_\beta\big\}\,(D^\beta B_{[6]}^\nu )\big)  + {\rm h.c.}  \,,
\label{del-L2S-L2V}
\end{eqnarray}
where further possible terms that are redundant owing to the  on-shell conditions 
of spin-$\frac32$ fields with $\gamma_\mu\,B_{[6]}^\mu = 0$ and $\partial_\mu\,B_{[6]}^\mu = 0$ are 
eliminated systematically. We note that the terms in (\ref{del-L2S-L2V}) imply contributions to the charmed baryon masses that do depend on the choice of the renormalization scale. Such terms need 
to be balanced by a set of symmetry breaking counter terms  that render the charmed baryon masses renormalization scale independent. 

We turn to the terms that break chiral symmetry explicitly. 
There are 7 symmetry breaking counter terms at order $Q^2$ and 16 terms of order $Q^4$. We recall from \cite{Lutz:2014jja,Heo:2018mur}
\allowdisplaybreaks[1]
\begin{eqnarray}
&&\mathcal{L}^{(2)}_\chi =  b_{1,[\bar 3\bar 3]}\,{\rm tr}\,\big( \bar B_{[\bar 3]}\,B_{[\bar 3]}\big)\,{\rm tr}\,\big(\chi_+\big)
+b_{2,[\bar 3\bar 3]}\,{\rm tr}\,\big( \bar B_{[\bar 3]}\,\chi_+\,B_{[\bar 3]}\big)
\nonumber\\
&& \qquad \;\,+\,b_{1,[\bar 3 6]}\,{\rm tr}\,\big( \bar B_{[6]}\,\chi_+\,B_{[\bar 3]} + {\rm h.c.}\big)
\nonumber\\
&& \qquad \;\,+\, 
b_{1,[66]}\,{\rm tr}\,\big( \bar B_{[6]}\,B_{[6]}\big)\,{\rm tr}\,\big(\chi_+\big)
+ b_{2,[66]}\,{\rm tr}\,\big( \bar B_{[6]}\,\chi_+\,B_{[6]}\,\big)
\nonumber\\
&& \qquad \;\,-\, d_{1,[66]}\,{\rm tr}\,\big(  g_{\mu \nu}\,\bar B_{[6]}^\mu\,B_{[6]}^\nu\big)\,{\rm tr}\,\big(\chi_+\big)
- d_{2,[66]}\,{\rm tr}\,\big( g_{\mu \nu}\,\bar B_{[6]}^\mu\,\chi_+\,B_{[6]}^\nu\,\big)\,,
\nonumber\\
&& \mathcal{L}_{\chi}^{(4)} \!= c_{1,[\bar 3\bar 3]}\,{\rm tr}\,\big( \bar B_{[\bar 3]}\,B_{[\bar 3]}\big)\,{\rm tr}\,\big(\chi_+^2\big)
+ c_{2,[\bar 3\bar 3]}\,{\rm tr}\,\big( \bar B_{[\bar 3]}\,B_{[\bar 3]}\big)\,\big({\rm tr}\,\chi_+\big)^2
\nonumber\\
&& \qquad \;+\, c_{3,[\bar 3\bar 3]}\,{\rm tr}\,\big( \bar B_{[\bar 3]}\,\chi_+\,B_{[\bar 3]}\,\big)\,{\rm tr}\,\big(\chi_+\big)
+  c_{4,[\bar3\bar3]}\,{\rm tr}\,\big( \bar B_{[\bar3]}\,\chi_+^2\,B_{[\bar3]}\big)
\nonumber\\
&& \qquad \; + \,c_{1,[66]}\,{\rm tr}\,\big( \bar B_{[6]}\,B_{[6]}\big)\,{\rm tr}\,\big(\chi_+^2\big)
+ c_{2,[66]}\,{\rm tr}\,\big( \bar B_{[6]}\,B_{[6]}\big)\,\big({\rm tr}\,\chi_+\big)^2
\nonumber\\
&& \qquad \;+\, c_{3,[66]}\,{\rm tr}\,\big( \bar B_{[6]}\,\chi_+\,B_{[6]}\,\big)\,{\rm tr}\,\big(\chi_+\big)
+ c_{4,[66]}\,{\rm tr}\,\big( \bar B_{[6]}\,\chi_+^2\, B_{[6]}\big)
\nonumber\\
&& \qquad \;+\,  c_{5,[66]}\,{\rm tr}\,\big( \bar B_{[6]}\,\chi_+\, B_{[6]}\,\chi^T_+\big)
\nonumber\\
&& \qquad \;+\, c_{1,[\bar 3 6]}\,{\rm tr}\,\big( \bar B_{[6]}\,\chi_+\,B_{[\bar 3]}+ {\rm h.c.}\,\big)\,{\rm tr}\,\big(\chi_+\big)
+  c_{2,[\bar3 6]}\,{\rm tr}\,\big( \bar B_{[6]}\,\chi_+^2\,B_{[\bar3]}+ {\rm h.c.}\big)
\nonumber\\
&& \qquad \; -\, e_{1,[66]}\,{\rm tr}\,\big( \bar B_{[6]}^\mu\,g_{\mu \nu}\,B_{[6]}^\nu\big)\,{\rm tr}\,\big(\chi_+^2\big)
- e_{2,[66]}\,{\rm tr}\,\big( \bar B_{[6]}^\mu\,g_{\mu \nu}\,B_{[6]}^\nu\big)\,\big({\rm tr}\,\chi_+\big)^2
\nonumber\\
&& \qquad \;-\, e_{3,[66]}\,{\rm tr}\,\big( \bar B_{[6]}^\mu\,g_{\mu \nu}\,\chi_+\,B_{[6]}^\nu\,\big)\,{\rm tr}\,\big(\chi_+\big)
- e_{4,[66]}\,{\rm tr}\,\big( \bar B_{[6]}^\mu\,g_{\mu \nu}\,\chi_+^2\, B_{[6]}^\nu\big)
\nonumber\\
&& \qquad \;-\, e_{5,[66]}\,{\rm tr}\,\big( \bar B_{[6]}^\mu\,g_{\mu \nu}\,\chi_+\, B_{[6]}^\nu\,\chi^T_+\big)\,.
\nonumber\\
&& \chi_+ = {\textstyle{1\over 2}}\, \big(
u \,\chi_0 \,u
+ u^\dagger \,\chi_0 \,u^\dagger \big)\,,
\label{def-L2-chi}
\end{eqnarray}
with $\chi_0 = 2\,B_0 \,{\rm diag }(m ,m, m_s)  $ proportional to the quark-mass matrix. We do not consider
isospin violating effects in this work. With $ \mathcal{L}_{\chi}^{(4)}$ 
the  16  symmetry breaking counter terms that  contribute to the charm baryon masses at  N$^3$LO  are shown. 

Altogether we count 54 low-energy constants in this section  that have to be determined by the data sets. Clearly, any additional constraints from 
heavy-quark spin-symmetry or large-$N_c$ QCD are desperately needed to arrive at any significant result. Such constraints were derived in \cite{Jenkins:1996de,Lutz:2014jja,Heo:2018mur} to subleading order in the $1/N_c$ expansion and  are summarized in Appendix A of \cite{Heo:2018qnk}. 
At subleading order there remain 23 independent low-energy constants only.

\section{Charmed baryon masses from QCD lattice simulations }

\begin{figure}[t]
\centering
\includegraphics[width=0.99\textwidth]{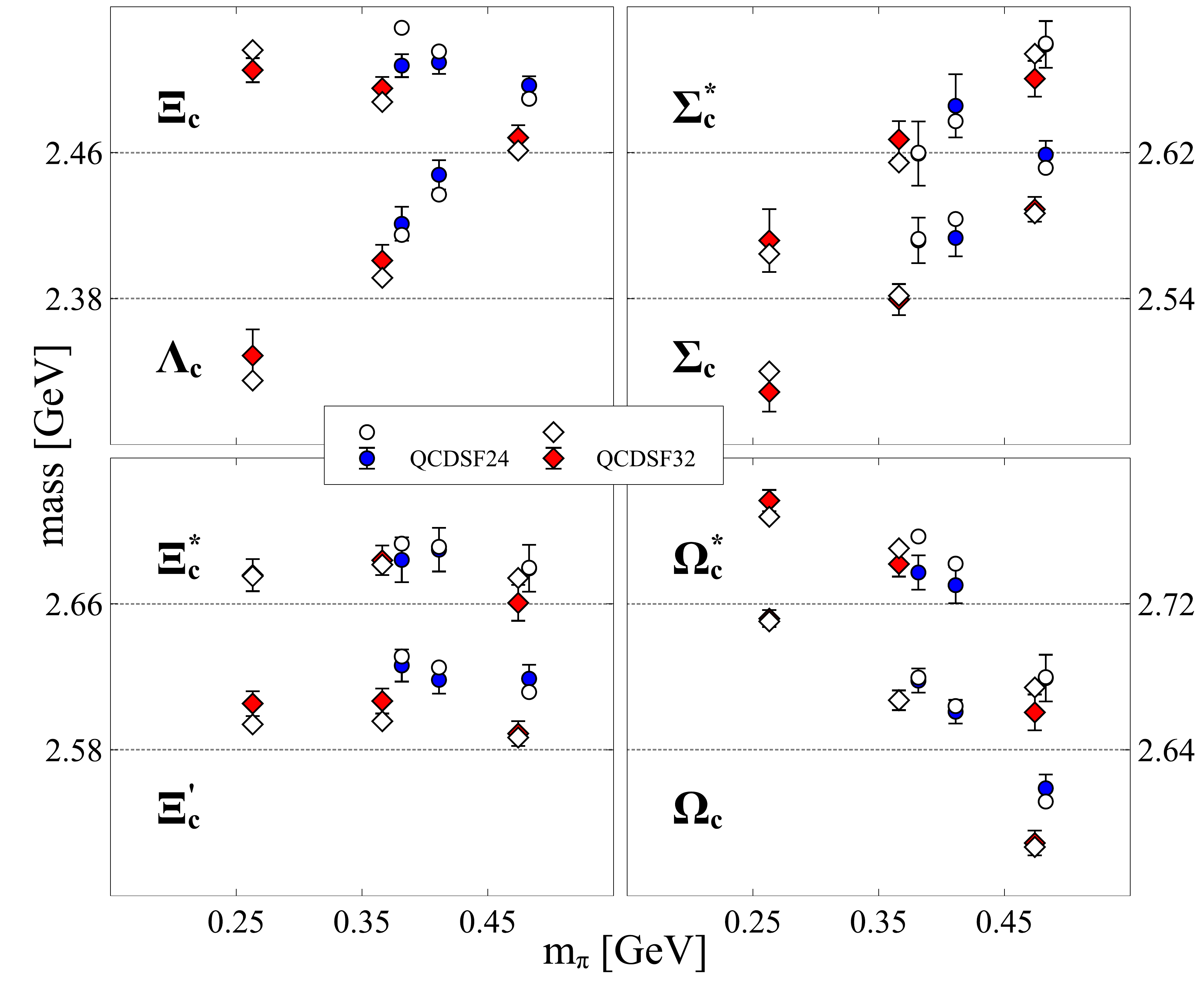}
\vskip-0.1cm
\caption{Our results from Fit 1 for the charmed baryon masses on the QCDSF-UKQCD ensembles \cite{Bali:2015lka}. The lattice results are given by blue ($24^3$ lattice) and red ($32^3$ lattice) filled symbols, where statistical errors are shown only. They are compared to the chiral extrapolation results in open symbols, which are always displayed on top of the lattice points.  }
\label{fig:1}
\end{figure}

We recall our strategy how to make use of the available QCD lattice data sets on hadron ground-state masses \cite{Lutz:2014oxa,Lutz:2018cqo,Guo:2018kno,Guo:2018zvl}. Altogether we consider 210 data points on lattice QCD ensembles with pion and kaon masses smaller than 600 MeV. To actually perform the fits is a computational challenge. For any set of the low-energy parameters ten coupled non-linear equations are to be solved on each lattice ensemble  
considered (see \cite{Heo:2018qnk}). We apply the evolutionary algorithm of GENEVA 1.9.0-GSI \cite{Geneva} with runs of a population size 2500 on 500 parallel CPU cores.

\begin{table}[t]
\setlength{\tabcolsep}{4.5mm}
\renewcommand{\arraystretch}{1.35}
\begin{center}
\begin{tabular}{l|rrr} 
                             &  Fit 1    &  Fit 2  &  Fit 3       \\  \hline


$M^{(1/2)}_{[\bar3]} $\hfill  [GeV]    & 2.468(${}_{-1}^{+0}$)     & 2.424(${}_{-3}^{+0}$)      &   2.415(${}_{-0}^{+0}$)   \\
$M^{(1/2)}_{[6]} $\hfill  [GeV]        & 2.515(${}_{-1}^{+1}$)     & 2.468(${}_{-4}^{+1}$)      &   2.388(${}_{-2}^{+4}$)   \\
$M^{(3/2)}_{[6]} $\hfill      [GeV]    & 2.594(${}_{-1}^{+1}$)     & 2.549(${}_{-3}^{+1}$)      &   2.586(${}_{-0}^{+0}$)   \\ \hline

$F_{[\bar36]} $                        & 0.7530                    & 0.7530                     & -0.7530   \\
$F_{[66]} $                            & 0.6502(${}_{-2}^{+6}$)    & 0.6500(${}_{-0}^{+5}$)     &  0.6745(${}_{-1}^{+2}$)   \\
$F_{[\bar3\bar3]} $                    & 0.0053(${}_{-54}^{+45}$)  & 0.0099(${}_{-05}^{+73}$)   & -0.0124(${}_{-1}^{+1}$)   \\ \hline 

$b_{1,[66]} $\hfill     [GeV$^{-1}$]   & 0.3372(${}_{-115}^{+101}$)&-0.2542(${}_{-145}^{+096}$) &  0.1053(${}_{-41}^{+46}$) \\
$b_{2,[66]} $\hfill     [GeV$^{-1}$]   &-0.3057(${}_{-198}^{+188}$)&-0.2048(${}_{-189}^{+377}$) & -0.5533(${}_{-79}^{+89}$) \\
$b_{2,[\bar3\bar3]}$\hfill [GeV$^{-1}$]&-0.5048(${}_{-198}^{+188}$)&-0.4048(${}_{-189}^{+377}$) & -0.7513(${}_{-79}^{+89}$) \\
$d_{2,[66]} $\hfill     [GeV$^{-1}$]   &-0.3330(${}_{-198}^{+188}$)&-0.2249(${}_{-179}^{+373}$) & -0.0624(${}_{-79}^{+89}$) \\ \hline \\

$10^3\,( 2\,L_6 -L_4 )\, $             & 0.0809(${}_{-18}^{+15}$)  & 0.0813(${}_{-16}^{+29}$)   &  0.0558(${}_{-2}^{+3}$)   \\
$10^3\,( 2\,L_8 -L_5)\, $              & 0.1098(${}_{-18}^{+18}$)  & 0.1498(${}_{-22}^{+02}$)   &  0.1500(${}_{-2}^{+0}$)   \\
$10^3\,(L_8 + 3\,L_7)\, $              & -0.5023(${}_{-7}^{+ 9}$)  & -0.5123(${}_{-12}^{+09}$)  & -0.5002(${}_{-1}^{+1}$)   \\
$m_s/ m $                              &  26.05(${}_{- 1}^{+ 1}$)  &   25.82(${}_{-0}^{+1}$)    &   25.79(${}_{- 0}^{+ 0}$) \\ 

\end{tabular}
\vskip-0.1cm
\caption{Results for LEC in (\ref{def-L1}, \ref{def-L2-chi}) based on three fit scenarios. Further LEC are implied by the NLO large-$N_c$ sum rules of \cite{Heo:2018mur}.
The low-energy constants $L_n$ are at the renormalization scale 
$\mu = 0.77$ GeV. We use $f = 92.4$ MeV throughout this work.}
\label{tab:Q2}
\end{center}
\end{table}

\begin{table}[t]
\setlength{\tabcolsep}{4.5mm}
\renewcommand{\arraystretch}{1.25}
\begin{center}
\begin{tabular}{l|rrr} 
                          &  Fit 1    &  Fit 2  &  Fit 3      \\  \hline

$c_{1,[66]} $\hfill    [GeV$^{-3}$]          & -0.5363(${}_{-128}^{+131}$) & -0.2786(${}_{-174}^{+280}$) & -0.3466(${}_{-45}^{+66}$)   \\
$c_{2,[66]} $\hfill    [GeV$^{-3}$]          &  0.0605(${}_{-74}^{+65}$)   &  0.1210(${}_{-55}^{+06}$)   & -0.0179(${}_{-31}^{+36}$)   \\
$c_{3,[66]} $\hfill    [GeV$^{-3}$]          & -0.3833(${}_{-60}^{+68}$)   & -0.3114(${}_{-085}^{+192}$) &  0.1812(${}_{-71}^{+34}$)   \\
$c_{4,[66]} $\hfill     [GeV$^{-3}$]         &  0.9850(${}_{-220}^{+150}$) &  0.8346(${}_{-682}^{+027}$) &  0.6941(${}_{-087}^{+134}$) \\
$c_{5,[66]} $\hfill     [GeV$^{-3}$]         & -0.3653(${}_{-20}^{+15}$)   & -0.3609(${}_{-42}^{+11}$)   & -0.4535(${}_{-07}^{+15}$)   \\

$c_{1,[\bar3\bar3]}$\hfill      [GeV$^{-3}$] & -0.7947(${}_{-101}^{+202}$) & -0.5741(${}_{-189}^{+484}$) & -0.9674(${}_{-45}^{+66}$)   \\
$c_{2,[\bar3\bar3]}$\hfill      [GeV$^{-3}$] &  0.1936(${}_{-74}^{+65}$)   &  0.2848(${}_{-158}^{+023}$) &  0.3489(${}_{-31}^{+36}$)   \\
$c_{3,[\bar3\bar3]}$\hfill      [GeV$^{-3}$] & -0.9100(${}_{-060}^{+146}$) & -0.8093(${}_{-217}^{+333}$) & -0.9759(${}_{-71}^{+34}$)   \\ 

$e_{1,[66]} $\hfill     [GeV$^{-3}$]         & -0.4842(${}_{-122}^{+136}$) & -0.2223(${}_{-174}^{+260}$) & -0.0643(${}_{-45}^{+66}$)   \\
$e_{3,[66]} $\hfill     [GeV$^{-3}$]         & -0.4048(${}_{-71}^{+75}$)   & -0.3347(${}_{-107}^{+160}$) & -0.2180(${}_{-60}^{+84}$)   \\
$e_{4,[66]} $\hfill     [GeV$^{-3}$]         &  0.9637(${}_{-210}^{+142}$) &  0.8012(${}_{-633}^{+065}$) &  0.0444(${}_{-087}^{+134}$) \\ \hline

$g^{(S)}_{0,[66]} $\hfill       [GeV$^{-1}$] & 0.5332(${}_{-216}^{+413}$) & -1.7697(${}_{-0860}^{+1538}$)& -1.3105(${}_{-28}^{+44}$)   \\
$g^{(S)}_{1,[66]} $\hfill       [GeV$^{-1}$] &0.2072(${}_{-0571}^{+1339}$)&  0.4406(${}_{-936}^{+925}$)  &  3.9737(${}_{-115}^{+567}$) \\
$g^{(V)}_{D,[66]} $\hfill       [GeV$^{-2}$] &0.4138(${}_{-1109}^{+0635}$)&  1.2902(${}_{-1794}^{+1282}$)& -2.9384(${}_{-316}^{+268}$) \\
$g^{(S)}_{0,[\bar3\bar3]}$\hfill [GeV$^{-1}$]& 1.0740(${}_{-377}^{+614}$) & -2.4883(${}_{-0670}^{+1758}$)&  2.6536(${}_{-158}^{+033}$) \\
$g^{(V)}_{0,[\bar3\bar3]}$\hfill [GeV$^{-2}$]& -1.7557(${}_{-388}^{+368}$)& -1.3281(${}_{-1111}^{+0551}$)& -5.2841(${}_{-38}^{+59}$)   \\

$h^{(S)}_{1,[66]} $\hfill       [GeV$^{-1}$] &-0.0716(${}_{-100}^{+030}$) & -0.3608(${}_{-545}^{+234}$)  & -0.4656(${}_{-5}^{+2}$)     \\
$h^{(S)}_{2,[66]} $\hfill       [GeV$^{-1}$] &1.0721(${}_{-1581}^{+1397}$)&  1.0279(${}_{-1699}^{+1312}$)& -0.0508(${}_{-703}^{+785}$) \\
$h^{(S)}_{4,[66]} $\hfill       [GeV$^{-1}$] &  2.3125(${}_{-168}^{+093}$)&  1.0016(${}_{-1478}^{+0143}$)&  6.1835(${}_{-33}^{+00}$)   \\
$h^{(S)}_{5,[66]} $\hfill       [GeV$^{-1}$] & -2.0747(${}_{-297}^{+525}$)& -0.2665(${}_{-0250}^{+2305}$)& -0.8207(${}_{-58}^{+67}$)   \\
$h^{(V)}_{1,[66]} $\hfill       [GeV$^{-2}$] & -0.5500(${}_{-552}^{+356}$)&  0.6440(${}_{-0335}^{+1568}$)& -6.7926(${}_{-076}^{+117}$) \\
$h^{(V)}_{2,[66]} $\hfill       [GeV$^{-2}$] & 0.2754(${}_{-261}^{+151}$) &  1.1689(${}_{-0603}^{+1255}$)&  0.8839(${}_{-41}^{+07}$)   \\

\end{tabular}
\vskip-0.1cm
\caption{Results for the LEC in (\ref{del-L2S-L2V}, \ref{def-L2-chi}) based on three fit scenarios. Further LEC are implied by the NLO large-$N_c$ sum rules of \cite{Heo:2018mur}. }
\label{tab:Q4}
\end{center}
\end{table}

\begin{figure}[t]
\centering
\includegraphics[width=0.99\textwidth]{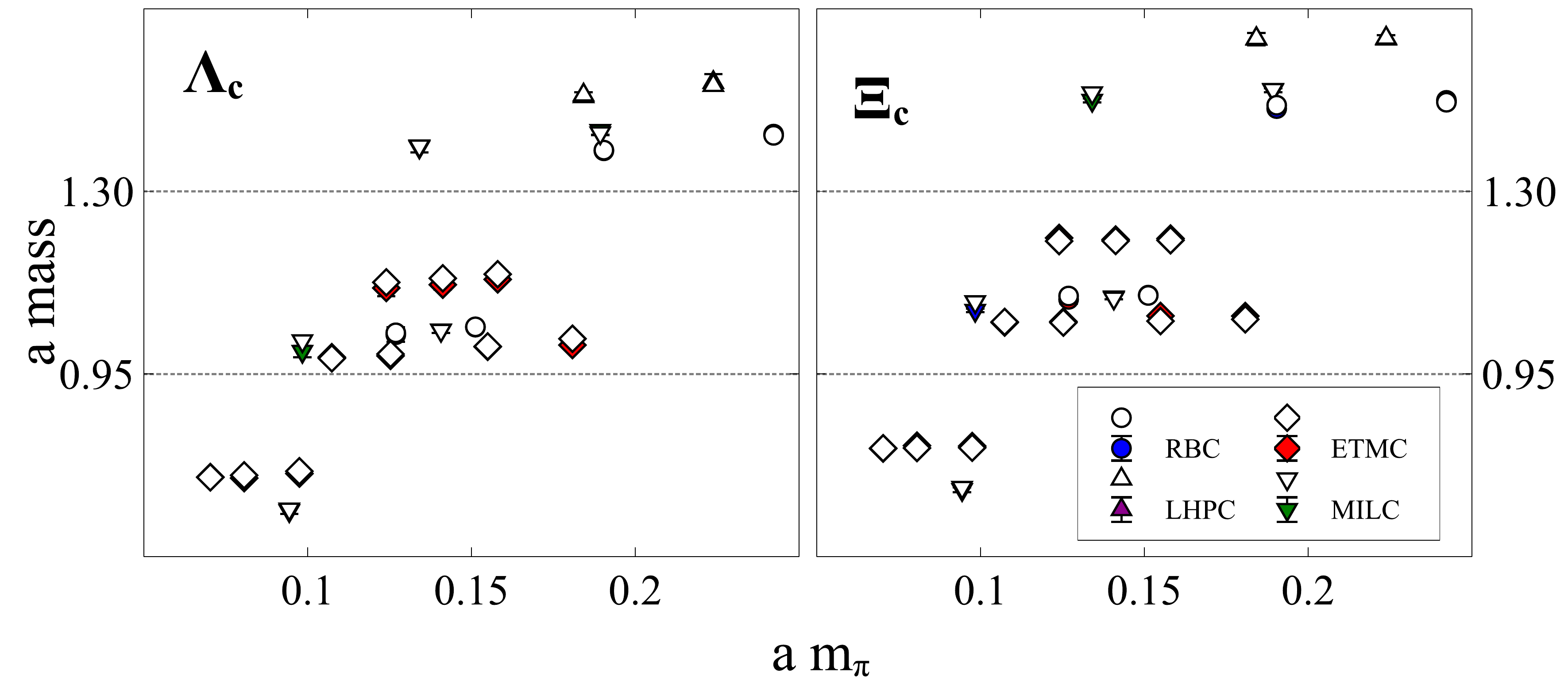}
\vskip-0.1cm
\caption{Our results Fit 1 for the charmed baryon masses on the 
 LHPC \cite{Liu:2009jc}, RBC-UKQCD \cite{Brown:2014ena}, MILC \cite{Briceno:2012wt} and ETMC \cite{Alexandrou:2014sha} ensembles. The extrapolation results in open symbols are always displayed on top of the colored lattice points. }
\label{fig:2}
\end{figure}

\begin{figure}[t]
\centering
\includegraphics[width=0.99\textwidth]{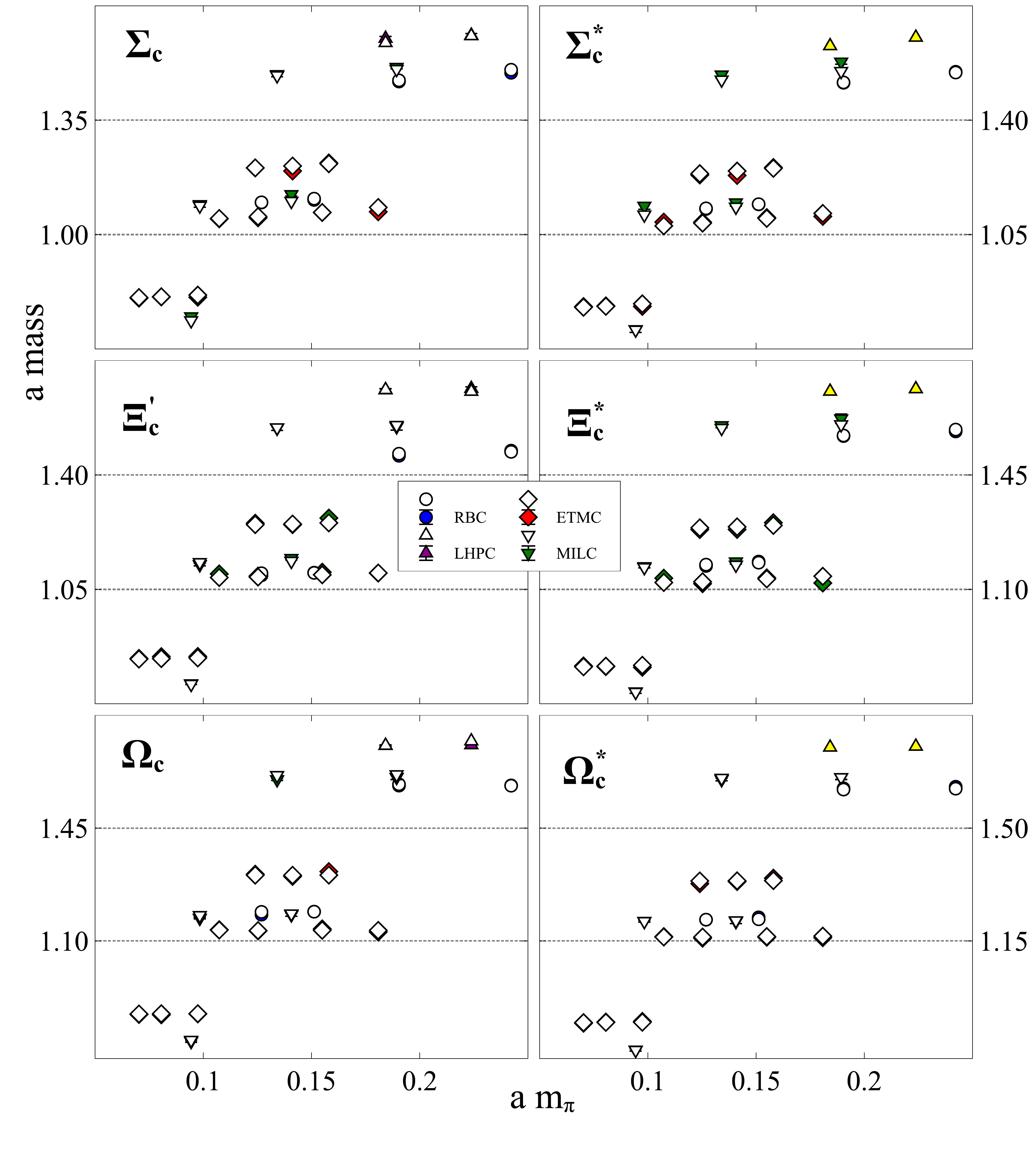}
\vskip-0.1cm
\caption{Continuation of Fig. 2. }
\label{fig:3}
\end{figure}

A subset of low-energy constants (LEC) is fixed by the requirement that the isospin averaged charmed baryon masses are reproduced as provided by the PDG \cite{PDG}. This amounts to a non-standard scale setting for the various lattice ensembles considered. For any given $\beta$ value characterizing a lattice ensemble we determine its associated lattice scale separately. From any such ensemble we take its published pion and kaon masses as given in lattice units and derive their associated quark masses. As in our previous works \cite{Lutz:2018cqo,Guo:2018kno} we do so at the one-loop level in terms of the low-energy constants of Gasser and Leutwyler. At the physical point the products $B_0\,m, B_0 \,m_s$ and $L_8 + 3\,L_7$ are set by the request to reproduce the empirical pion, kaon and eta masses. That leaves  two particular combinations $2\, L_6- L_4 $ and $  2\,L_8 -L_5 $ undetermined only. As emphasized repeatedly the lattice data sets on   hadron masses are quite sensitive to the latter and permit an accurate determination of their values \cite{Lutz:2018cqo,Guo:2018kno}. Consistent results were obtained from two independent analysis based on distinct lattice data sets on the hadron masses. This suggests the ranges 
$10^3\,(2\, L_6 -L_4) = 0.04 - 0.16$ and $ 10^3\,(2\,L_8 - L_5) =0.02 -0.11$ at the renormalization scale $\mu =0.77$ GeV. In Tab. \ref{tab:Q2}-\ref{tab:Q4} we collect the values of the LEC according to three fit scenarios. Sets of independent LEC are shown only. The remaining ones follow from 
the NLO large-$N_c$ sum rules of \cite{Heo:2018mur,Heo:2018qnk}.
For instance we recall
\begin{eqnarray}
&&  C_{[66]} = \frac{\sqrt 3}{2}\,\,\big( F_{[\bar 3\bar 3]}- F_{[66]} \big) \,, \qquad \qquad \;\;\;\,C_{[\bar 36]} = {\sqrt 3}\,F_{[\bar 36]}  \,,
\nonumber\\
&& H_{[66]} = \frac{3}{2}\,\,\big( F_{[\bar 3\bar 3]}+ F_{[66]} \big) \,, \qquad \qquad \qquad
b_{1,[66]} = d_{1,[66]} = b_{1,[\bar{3}\bar{3}]}
\nonumber\\
&&   b_{1,[\bar{3}6]} = \frac{1 }{\sqrt{3} }\,\big(b_{2,[66]} - d_{2,[66]}\big) \,.
\end{eqnarray}
Our Fit 1 is our most reasonable scenario with an excellent reproduction of the lattice data sets. In  

Fig.~\ref{fig:1} we confront its implications for the baryon masses as computed by the QCDSF-UKQCD  collaboration \cite{Bali:2015lka}.  Further results are shown in Fig. \ref{fig:2} and Fig. \ref{fig:3} for the baryon masses on LHPC \cite{Liu:2009jc}, RBC-UKQCD \cite{Brown:2014ena}, MILC \cite{Briceno:2012wt} and ETMC \cite{Alexandrou:2014sha} ensembles.

Such lattice data are well reproduced. Similar results are observed for the Fit 2 scenario which 
rests on a slightly different way how the systematic uncertainties in our approach are dealt with. In Fit 3 we explore a parameter set which does not quite match the success of Fit 1 and Fit 2 
in the reproduction of the lattice data. Fit 3 is based on a negative value of $F_{[\bar 3 6]}= -0.753$. While the magnitude  $|F_{[\bar 3 6]}|= 0.753$ was estimated from empirical values of some hadronic decay widths parameters \cite{Heo:2018qnk} its sign (in our phase convention as used in the derivation of the large-$N_c$ sum rules) remained undetermined. From a recent lattice computation of axial charges of some charmed baryons \cite{Alexandrou:2016xok} we may estimate $F_{[6 6]} \sim 0.7 $, where in this case the computation does determine the sign of $F_{[66]}$ unambiguously. The condition $F_{[66]}> 0.65$ is imposed throughout our analysis.
We wish to emphasize an important finding. For the LEC in Fit 1 and Fit 2 we observe that the 
hierarchy  of the large-$N_c$ sum rules is well followed. For instance at LO we expect 
$F_{[ \bar 3\bar 3]} = 0$, $b_{1,[\bar 36]} =c_{n,[\bar 36]} = 0$ and $c_{n,[66]} =e_{n,[66]} $. Further such relations can be traced in \cite{Heo:2018qnk}.

\begin{table}[t]
\setlength{\tabcolsep}{1.4mm}
\renewcommand{\arraystretch}{1.15}
\begin{center}
\begin{tabular}{l|rrr|l} 
                                                                     &  Fit 1    &  Fit 2   &  Fit 3  &  lattice \\ \hline

$a_{\rm QCDSF-UKQCD}\,  \hfill \mathrm{[fm]}$                        &  0.0727(${}_{-0}^{+1}$)  & 0.0734(${}_{- 4}^{+ 1}$) &  0.0738(${}_{-1}^{+1}$)  & 0.072(4)\cite{Bali:2015lka} \\
$a\,\Delta_{c,\rm QCDSF-UKQCD}$                                      & -0.0198(${}_{-2}^{+9}$)  &-0.0064(${}_{-47}^{+09}$) & -0.0032(${}_{-4}^{+6}$)  & \\
$\chi^2/N$                                                           & 0.40                     & 0.63                     & 0.62                     & \\ \hline

$a_{\rm LHPC} \,   \hfill \mathrm{[fm]}$                             & 0.1282(${}_{- 1}^{+ 1}$) & 0.1273(${}_{- 4}^{+ 1}$) & 0.1276(${}_{-4}^{+2}$)   & 0.1243(25)\cite{Liu:2009jc}\\ 
$a\,\Delta_{c,\rm LHPC}$                                             & 0.0200(${}_{-10}^{+00}$) & 0.0200(${}_{-34}^{+00}$) &-0.0200(${}_{-32}^{+00}$) & \\
$\chi^2/N$                                                           & 0.28                     & 0.75                     & 0.68                     & \\ \hline

$a^{\beta \simeq 6.76}_{c,\rm RBC-UKQCD}\,   \hfill \mathrm{[fm]}$   & 0.1179(${}_{- 1}^{+ 1}$) & 0.1186(${}_{- 2}^{+ 0}$) & 0.1186(${}_{-1}^{+0}$)   & 0.1119(17)\cite{Brown:2014ena}\\ 
$a\,\Delta^{\beta \simeq 6.76}_{c,\rm RBC-UKQCD}$                    & 0.0320(${}_{-36}^{+22}$) & 0.0493(${}_{-83}^{+88}$) & 0.0443(${}_{-14}^{+65}$) & \\
$\chi^2/N$                                                           & 0.42                     & 0.78                     & 0.80                     & \\ \hline

$a^{\beta \simeq 7.09}_{c,\rm RBC-UKQCD}\,   \hfill \mathrm{[fm]}$   & 0.0885(${}_{- 1}^{+ 0}$) & 0.0890(${}_{- 2}^{+ 1}$) & 0.0888(${}_{-1}^{+1}$)   & 0.0849(17)\cite{Brown:2014ena}\\ 
$a\,\Delta^{\beta \simeq 7.09}_{c,\rm RBC-UKQCD}$                    & 0.0267(${}_{-26}^{+14}$) & 0.0384(${}_{-61}^{+52}$) & 0.0335(${}_{-09}^{+58}$) & \\
$\chi^2/N$                                                           & 0.48                     & 1.00                     & 0.61                     & \\
 
\end{tabular}
\vskip-0.1cm
\caption{Specifics for Fit 1 - 3. The offset parameter $a\,\Delta_c$ is introduced in (\ref{def-Deltac}). While Fit 1 and Fit 3 are based on a systematical error ansatz of 10 MeV, Fit 2 rests on 5 MeV.}
\label{tab:lattice-scale-A}
\end{center}
\end{table}

\begin{table}[t]
\setlength{\tabcolsep}{2.5mm}
\renewcommand{\arraystretch}{1.15}
\begin{center}
\begin{tabular}{l|rrr|l}  
                                                                 &  Fit 1      &  Fit 2  &  Fit 3  &     lattice  \\ \hline

$a^{\beta = 1.90 }_{c,\rm MILC}\,   \hfill \mathrm{[fm]}$        & 0.1234(${}_{-3}^{+2}$)  & 0.1210(${}_{-2}^{+4}$)    & 0.1230(${}_{-1}^{+1}$)  &  0.1193(9)\cite{Briceno:2012wt}\\
$a\,\Delta^{\beta = 1.90}_{c,\rm MILC}$                          & 0.0493(${}_{-70}^{+30}$)& 0.0285(${}_{-166}^{+186}$)& 0.0461(${}_{-8}^{+4}$)  & \\
$\chi^2/N$                                                       & 1.41                    & 2.11                      & 0.78                    & \\ \hline

$a^{\beta = 1.95}_{\rm MILC}\,   \hfill \mathrm{[fm]}$           & 0.0903 (${}_{-4}^{+1}$) & 0.0885(${}_{-1}^{+5}$)    & 0.0904(${}_{-0}^{+0}$)  &  0.0871(11)\cite{Briceno:2012wt}\\
$a\,\Delta^{\beta =1.95 }_{c,\rm MILC}$                          & 0.0295(${}_{-47}^{+17}$)& 0.0147(${}_{-093}^{+134}$)& 0.0332(${}_{-6}^{+3}$)  & \\
$\chi^2/N$                                                       & 1.96                    & 2.90                      & 1.50                    & \\\hline

$a^{\beta = 2.10}_{\rm MILC}\,   \hfill \mathrm{[fm]}$           & 0.0603(${}_{-3}^{+1}$)  & 0.0592(${}_{-1}^{+4}$)    & 0.0607(${}_{-2}^{+0}$)  & 0.0578(23)\cite{Briceno:2012wt}\\
$a\,\Delta^{\beta = 2.10}_{c,\rm MILC}$                          & 0.0157(${}_{-29}^{+09}$)&0.0061(${}_{-45}^{+89}$)   & 0.0218(${}_{-3}^{+2}$)  & \\
$\chi^2/N$                                                       & 0.44                    & 0.64                      & 0.47                    & \\\hline

$a^{\beta = 1.90 }_{c,\rm ETMC}\,   \hfill \mathrm{[fm]}$        & 0.0956(${}_{-1}^{+1}$)  & 0.0959(${}_{-4}^{+2}$)    & 0.0965(${}_{-1}^{+1}$)  & 0.0936(13)\cite{Alexandrou:2014sha}\\
$a\,\Delta^{\beta = 1.90}_{c,\rm ETMC}$                          &-0.0497(${}_{-11}^{+11}$)&-0.0400(${}_{-50}^{+11}$)  &-0.0347(${}_{-09}^{+11}$)& \\ 
$\chi^2/N$                                                       &  1.06                   & 1.50                      & 1.19                    & \\\hline

$a^{\beta = 1.95}_{\rm ETMC}\,   \hfill \mathrm{[fm]}$           & 0.0832(${}_{-0}^{+1}$)  & 0.0834(${}_{-3}^{+1}$)    & 0.0838(${}_{-0}^{+0}$)  & 0.0823(10)\cite{Alexandrou:2014sha}\\
$a\,\Delta^{\beta =1.95 }_{c,\rm ETMC}$                          &-0.0406(${}_{-9}^{+9}$)  &-0.0331(${}_{-43}^{+09}$)  & -0.0301(${}_{-8}^{+9}$) & \\ 
$\chi^2/N$                                                       & 1.18                    & 1.70                      & 1.46                    & \\\hline

$a^{\beta = 2.10}_{\rm ETMC}\,   \hfill \mathrm{[fm]}$           & 0.0643(${}_{-1}^{+1}$)  & 0.0644(${}_{-2}^{+1}$)    & 0.0646(${}_{-1}^{+0}$)  & 0.0646(7)\cite{Alexandrou:2014sha}\\
$a\,\Delta^{\beta = 2.10}_{c,\rm ETMC}$                          &-0.0283(${}_{-7}^{+7}$)  &-0.0236(${}_{-32}^{+06}$)  & -0.0232(${}_{-6}^{+6}$) & \\ 
$\chi^2/N$                                                       & 0.53                    & 0.78                      & 0.74                    & \\
 
\end{tabular}
\vskip-0.1cm
\caption{Continuation of Tab. \ref{tab:lattice-scale-A}. }
\label{tab:lattice-scale-B}
\end{center}
\end{table}

Since we do not consider discretization effects and also have a residual uncertainty in our one-loop chiral extrapolation approach we assigned each baryon mass a systematical error that is added to its statistical error in quadrature in our various fits. Our chisquare function, $\chi^2$, assumes a universal systematical error for the charmed baryon masses. Its size of 10 MeV was chosen to arrive at about $\chi^2/N \simeq 1$ in Fit 1, with $N$ the number of fitted baryon masses. In contrast Fit 2 rests on a systematic error ansatz of 5 MeV instead.
In addition 
we introduced a  universal parameter $\Delta_c$  of the form
\begin{eqnarray}
a\,M_H \to a\, M_H + (1 + \epsilon_H )\,a\,\Delta_c\,, \qquad {\rm with }\qquad \epsilon_H \simeq 0\,,
\label{def-Deltac}
\end{eqnarray}
with which  the choice of the charm quark mass can be fine tuned. The values of $\epsilon_H$ depend not only on the light quark masses and the type of charmed baryon considered,  but also on the $\beta_{\rm QCD} $ value of the ensemble considered. Given the charmed baryon masses at two distinct 
values of the charm quark mass $\Delta_c$ and $\epsilon_H$ can be determined \cite{Guo:2018kno}.
We do so for the LHPC and MILC ensembles. For the other ensembles 
such data are not available in the literature and therefore we can access the parameters $\Delta_c$ only, i.e. we assume $\epsilon_H =0$ in this case. 
Non-vanishing values for $\Delta_c$ hint at a possible mismatch of the chosen charm-quark mass in a given lattice ensemble or a leading-order discretization effect as it may be implied by our non-standard scale setting approach.

In Tab. \ref{tab:lattice-scale-A} we present the results for such offset parameters on ensembles of QCDSF-UKQCD, LHPC and RBC-UKQCD as they follow in our fit scenarios. 
In addition the lattice scales as well as the quality of the data reproduction are quantified. 
We find significant sizes for the offset parameters. Note, however, that from those results we can not infer a mismatch of the chosen charm quark mass. This is so since we can not exclude the presence of discretization effects which may be parameterized by a contribution to $\Delta_c \sim a$ proportional to the lattice scale. A minimal model to discriminate an offset in the charm quark mass from the presence of discretization effects is
\begin{eqnarray}
a\,M_H \to a\, M_H +\,a\,\bar \Delta_c +  a^2\, \bar \Delta_d\,,
\label{def-Deltad}
\end{eqnarray}
where the parameters $\bar \Delta_c$ and $\bar \Delta_d$ should not depend on the chosen ensemble for a given lattice setup. All results of our current work are based on the ansatz (\ref{def-Deltad}). We find $\bar \Delta^{\rm RBC-UKQCD}_c = 77^{+1}_{-5}$ MeV and $95^{+2}_{-13}$ MeV from Fit 1 and Fit 2 respectively. 
We turn to the ensembles of ETMC and MILC, both of which are available at three distinct beta values. In Tab. \ref{tab:lattice-scale-B} our results for the offset parameters, lattice scales and chisquare values are presented. From Fit 1 we find 
$\bar \Delta^{\rm MILC}_c =  25^{+1}_{-8}$ MeV and $\bar \Delta^{\rm ETMC}_c = -55^{+2}_{-2} $ MeV.
Corresponding values from Fit 2  are $-5^{+29}_{-4}$ MeV and $-52^{+1}_{-9}$ MeV.

\begin{figure}[t]
\centering
\includegraphics[width=0.99\textwidth]{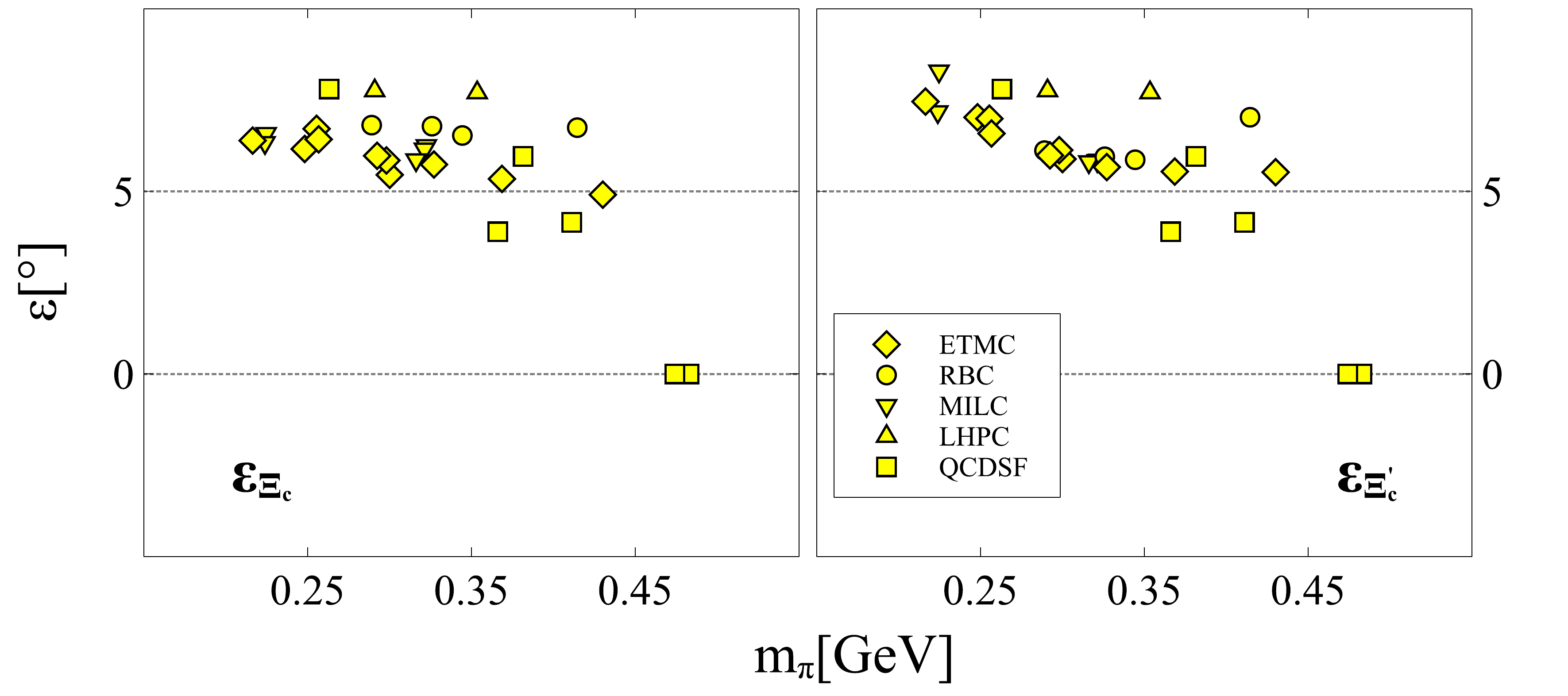}
\vskip-0.1cm
\caption{The $\Xi_c-\Xi_c'$ mixing angles on the considered lattice ensembles.  }
\label{fig:4}
\end{figure}

In Fig. \ref{fig:4} we show the $\Xi_c-\Xi_c'$ mixing angles evaluated 
on all considered lattice ensembles in Fit 1. Given our approach the mixing angle acquires 
a mass dependence, and therefore two distinct values are implied. 
At the physical point the mixing angles reach $\epsilon_\Xi \simeq 0.103(_{-4}^{+3})$ and  $\epsilon_{\Xi_c'} \simeq  0.117(_{-2}^{+2})$ at the $\Xi_c$ and $\Xi_c'$ masses respectively. We find rather small values for 
the mixing angles on all considered lattice ensembles.

We wish to comment on the uncertainties in the LEC of Tab. \ref{tab:Q2}-\ref{tab:Q4}. We derived the one-sigma statistical errors. Corresponding errors in Tab. \ref{tab:lattice-scale-A}-\ref{tab:lattice-scale-B} are derived for the various lattice scales and offset parameters.
Since a quite large set of data points is fitted in terms of a significantly smaller number of LEC,  any derived statistical error is of almost no physical relevance. Uncertainties are largely dominated by the systematic uncertainties, which can be made quantitative only after a better control of discretization effects and the generic form of two-loop contributions are available.

\section{Summary and outlook}

In this Letter we documented a first chiral extrapolation fit to the world data 
on charmed baryon masses on QCD lattices. The set of low-energy constants that determine 
the baryon masses at N$^3$LO was obtained. Given the latter the way is paved to explore 
the coupled-channel systems of Goldstone bosons and charmed baryon ground states. Such systems are of particular importance in QCD since, like in the open-charm meson sector, flavour exotic multiplet are expected here.

\vskip0.2cm
\centerline{\bf{Acknowledgments}}
\vskip0.2cm
Constantia Alexandrou, Gunnar Bali, John Bulava, Utku Can and Daniel Mohler are acknowledged for stimulating discussions. Y. H. received partial support from Suranaree University of Technology, Office of the Higher Education Commission under NRU project of Thailand (SUT-COE: High Energy Physics and Astrophysics) and SUT-CHE-NRU (Grant No. FtR.11/2561).

\bibliographystyle{elsarticle-num}
\bibliography{literature}

\end{document}